\documentclass[fleqn,usenatbib]{mnras}
\usepackage{newtxtext,newtxmath}
\usepackage{booktabs}
\usepackage[T1]{fontenc}

\DeclareRobustCommand{\VAN}[3]{#2}
\let\VANthebibliography\thebibliography
\def\thebibliography{\DeclareRobustCommand{\VAN}[3]{##3}\VANthebibliography}

\usepackage{graphicx}	
\usepackage{amsmath}	

\title[AGN and Star Formation in NGC\,232]{AGN feedback and star formation in the peculiar galaxy NGC\,232:  Insights from VLT-MUSE Observations}
\author[J. H. Costa-Souza et al.]{Jos\'e Henrique Costa-Souza,$^{1}$\thanks{E-mail: henrique.souza@acad.ufsm.br  (JHCPS)} Rogemar A. Riffel,$^{1}$\thanks{E-mail: rogemar@ufsm.br  (RAR)} Oli L. Dors,$^{2}$ Rog\'erio Riffel,$^{3,4}$ 
\newauthor Paulo C. da Rocha-Poppe$^{5,6}$
 \\
$^{1}$Departamento de F\'isica, CCNE, Universidade Federal de Santa Maria, Av. Roraima 1000, 97105-900, Santa Maria, RS, Brazil\\
$^{2}$Universidade do Vale do Para\'iba, Av. Shishima Hifumi, 2911, 12244-000, S\~ao Jos\'e dos Campos, SP, Brazil \\
$^{3}$ Departamento de Astronomia, Instituto de F\'\i sica, Universidade Federal do Rio Grande do Sul, CP 15051, 91501-970, Porto Alegre, RS, Brazil \\
$^{4}$Instituto de Astrof\'\i sica de Canarias, Calle V\'\i a L\'actea s/n, E-38205 La Laguna, Tenerife, Spain\\
$^{5}$Departamento de F\'isica, Universidade Estadual de Feira de Santana, Av. Transnordestina S/N, Novo Horizonte, CEP 44036-900, Feira de Santana, BA, Brazil\\
$^{6}$UEFS, Observatório Astronômico Antares, Rua da Barra 925, Jardim Cruzeiro, CEP 44024-432,
Feira de Santana, BA, Brazil\\
}
\date{Accepted XXX. Received YYY; in original form ZZZ}

\pubyear{2015}

\begin{document}
\label{firstpage}
\pagerange{\pageref{firstpage}--\pageref{lastpage}}
\maketitle

\begin{abstract}
We use VLT-MUSE integral field unit data to study the ionized gas physical properties and kinematics as well as the stellar populations of the Seyfert 2 galaxy NGC\,232 as an opportunity to understand the role of AGN feedback on star formation. The data cover a field of view of 60$\times$60 arcsec$^{2}$ at a spatial resolution of $\sim$\,850\,pc. The emission-line profiles have been fitted with two Gaussian components, one associated to the emission of the gas in the disc and the other due to a bi-conical outflow. The spectral synthesis suggests a predominantly old stellar population with ages exceeding 2\,Gyr, with the largest contributions seen at the nucleus and decreasing outwards. Meanwhile, the young and intermediate age stellar populations exhibit a positive gradient with increasing radius and a circum-nuclear star forming ring with radius of $\sim$0.5\,kpc traced by stars younger than 20 Myr, is observed. This, along with the fact that AGN and SF dominated regions present similar gaseous oxygen abundances, suggests a shared reservoir feeding both star formation and the AGN. We have estimated a maximum outflow rate in ionized gas of $\sim$1.26\,M${\odot}$\,yr$^{-1}$ observed at a distance of $\sim$560 pc from the nucleus. The corresponding maximum kinetic power of the outflow is $\sim3.4\times10^{41}$ erg\,s$^{-1}$. This released energy could be sufficient to suppress star formation within the ionization cone, as evidenced by the lower star formation rates observed in this region.
\end{abstract}




\section{Introduction}

In the hierarchical scenario of galaxy formation, galaxy interactions and mergers play a fundamental role and they are the key to the growth of the structures observed in the local universe. Galaxy interactions can modify the morphological, physical \citep{1966apg..book.....A,1987cspg.book.....A,2014MNRAS.437.1155K} and chemical properties \citep{2008MNRAS.389.1593K,2011MNRAS.417..580P,2014MNRAS.444.2005R,2023ApJ...952...77O} of the involved objects. Galaxy mergers can trigger the star formation of a galaxy, causing extreme bursts of star formation that can last a few million years \citep{2002MNRAS.333..327T,2008A&A...492...31D,2018AJ....156..295P}. A major merger can destabilize large amounts of gas, causing it to stream towards the center and providing a gas reservoir that can fuel not only circumnuclear star formation but also feed the supermassive black hole (SMBH), triggering an Active Galactic Nucleus (AGN, e.g. \citet{2000ApJ...544..747S,2001ApJ...546..845G,sb19}).  An AGN-starburst connection can also occur when the gas reservoir initially fuels circumnuclear star formation, and the subsequent mass loss from evolving stars triggers the nuclear activity  \citep{norman88,1997ApJ...482..114H,1995ApJ...448...98H,1998ApJ...495..698G,rogemar_09}. Indeed, observational studies have shown that the most luminous AGNs are observed in galaxy mergers \citep{Urrutia08,treister12,trakhtenbrot17}. 

 In addition to external influences, internal processes like feedback from an AGN can impact the evolution of the host. AGN feedback can manifest either heating, or blowing the cold gas out of it. Otherwise compressing or redistributing the gas, resulting in changes at the star formation in the host.  Either way the impact AGN feedback is invoked to be a essential ingredient of galaxy evolution models \citep{dimatteo05,hopkins10,harrison17,harrison18}. Since both AGNs and nuclear star formation can be originate from merger-induced gas inflows, the study of AGNs hosted by interacting/peculiar galaxies offers a unique opportunity to investigate the connection, or its absence, between AGN and starburst activities \citep{Perry85,Terlevich85,norman88}.

It is widely known that once the AGN has been triggered, particles orbiting the supermassive black hole (SMBH) can reach velocities near the speed of light, heating the surrounding matter through friction until it approaches the Eddington limit \citep{2005MNRAS.363L..91C,2014ARA&A..52..529Y}, at which point the radiation pressure becomes sufficiently high to expel matter from the SMBH accretion disc, leading to nuclear gas outflows. These outflows can redistribute the gas within the galaxy and even eject it from the system, thereby significantly impacting the star formation within the host galaxy \citep{2014ARA&A..52..529Y}.  In some cases, the nuclear winds can compress the Interstellar Medium (ISM) and induce star formation, resulting in a "positive" AGN feedback process \citep{2017Natur.544..202M,2019MNRAS.485.3409G}. On the other hand, if the radiation pressure or the turbulence caused by AGN winds prevents the gas from cooling and collapsing, it suppresses star formation, implying a negative AGN feedback \citep{harrison17,2005MNRAS.361..776S}. The effect of the energy released by AGN radiation and winds on the host galaxy is often referred to as the quasar or radiative feedback mode \citep{2008ARA&A..46..475H,2018NatAs...2..198H}. 

In addition, galaxy interactions and mergers can lead to the flow of gas with low metallicity from the outer parts of a galaxy's disc to the central part. This process decreases the metallicity in the inner regions and modifies the radial abundance gradients \citep{2010ApJ...721L..48K, 2014MNRAS.444.2005R,2008MNRAS.389.1593K}.  Gas phase metalicity also seems to anti-correlate with the X-ray luminosity \citep{2023MNRAS.520.1687A}, which is another indication that cold gas inflows might be the main mechanism of gas impoverishment at the central regions.  Recent studies \citep{2022MNRAS.513..807D} indicate that AGNs host a metal-poor gas reservoir, which could potentially have originated from past interactions. On the other hand, AGN winds can redistribute gas in galaxies and detailed studies of the gas kinematics are necessary to properly understand the role of AGN feedback in galaxy evolution.


Here, we use integral field spectroscopy to map the gas emission structure in the AGN host NGC\,232. This galaxy is located in the the local universe at a distance of $95.78\pm6.74$ Mpc \citep{2009MNRAS.399..683J}, and is often classified as an Ultra Luminous Infrared Galaxy (ULIRG), and in some cases, even as an Extremely Luminous Far-infrared galaxy \citep[ELF; ]{1990ApJ...359..291N}. 
NGC\,232 forms a galaxy pair along with NGC\,235 and it has been subject of numerous studies e.g. analyzing the influence of close pairs to the AGN triggering  by comparing the kinematic and physical properties of both objects, as well as an assessment of the presence of tidal features \citep{1996AJ....111..696K}; also, a study of the overall star formation on interacting systems \citep{2002ApJS..143...47D} verified a trail of H$\alpha$ emission between the pair, seen for the first time in \citep{1994AJ....107...99R}. NGC\,232 features an unusual highly ionized optical jet-like structure, as traced by the [O\,{\sc iii}]$\lambda$5007 emission, extending up to  $\sim$2.7\,kpc from the nucleus \citep{2017ApJ...850L..17L}. This structure is the second largest optical jet seen in AGN, being only smaller than the jet seen in 3C\,120.  

The object is also part of the Great Observatories All-Sky LIRG Survey (GOALS, \citet{2009PASP..121..559A}) sample, and SMA B0DEGA (SMA Below 0 DEgree GAlaxies, \citet{2010gama.conf...97E}) project, being one of the most IR luminous galaxies of the last one. This led to a dedicated study for the CO(2–1) gas, in which was noted asymmetric wings over the velocity map aligned along the PA$\,=\,44^{\circ}$ \citep{2018ApJ...866...77E}.

In this work, we present a detailed study of gas emission structure and properties of the stellar populations of NGC\,232 using the same data presented by \citet{2017ApJ...850L..17L}, obtained with the Multi Unit Spectroscopic Explorer (MUSE) on the Very Large Telescope (VLT). The remarkable spatial coverage and resolution of MUSE, allow us to further explore the aspects of the optical-jet regarding metal abundances, source of emission, and also star formation history. This work is organized as follows: Section~\ref{sec:obs} presents the data and the analysis methods. The results on the stellar populations and emission gas properties are shown in Sec.~\ref{sec:results} and discussed in Sec.~\ref{sec:disc}. Finally, Sec.~\ref{sec:conc} presents our conclusions.

\section{Observations and Methods}\label{sec:obs}

\subsection{The data}
The data we explored is originated from the All-Weather MUse Supernova Integral field Nearby Galaxies (AMUSING) project \citep{amusing}. The data was originally acquired for the study of the supernova Ia SN2006et, using the Multi Unit Spectroscopic Explorer (MUSE) instrument on the VLT, operated by the European Southern Observatory (ESO). The observations were done on June 27, 2015 with a total  on source exposure time of 555 sec.   We use the data cube of NGC\,232 provided by the ESO Science Portal\footnote{\url{https://archive.eso.org/scienceportal/home}}, which covers a field of view of $\sim 60\times 60$\, arcsec$^2$, with an angular sampling of  $\sim 0.2\times0.2$ arcsec$^2$ per spaxel. The spectral coverage is 4750--9350\AA, with a spectral dispersion of 1.25\AA.  The seeing during the observations of $0\farcs77\,\pm\,0\farcs 07$ corresponds to a spatial resolution of $\sim$375\,pc at the galaxy.

\begin{figure*}
    \centering
    \includegraphics[width=0.98\textwidth]{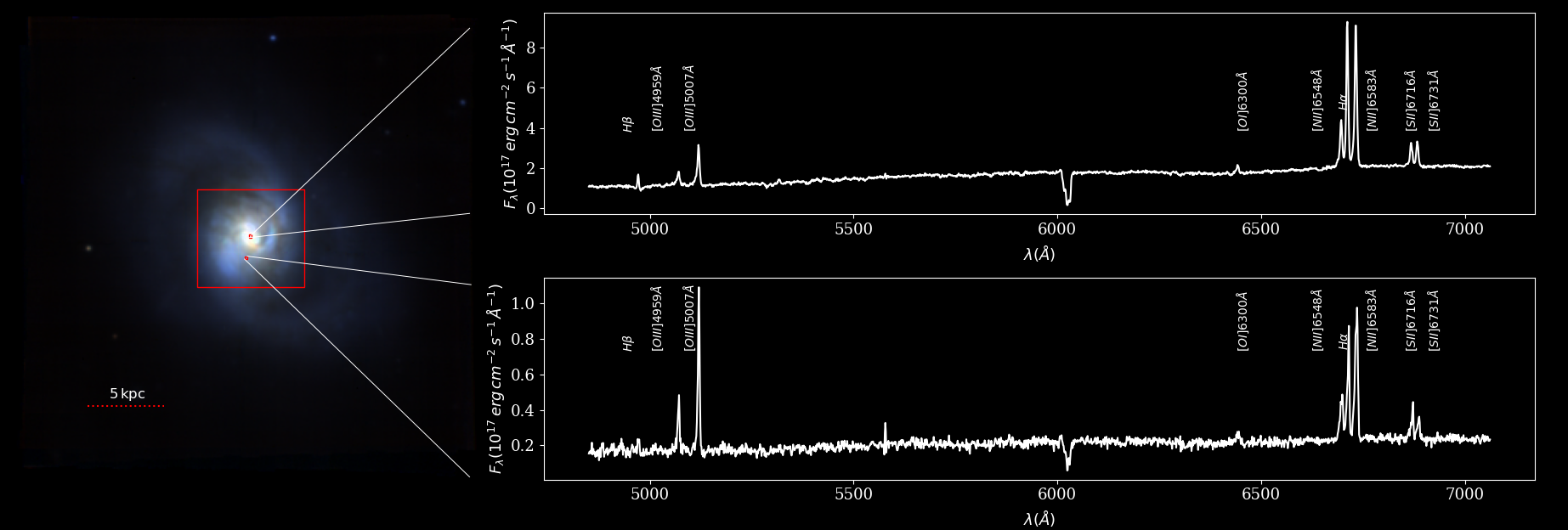}
    \caption{RGB composite image of NGC\,232, made using the the MUSE data cube. The red squares indicates the positions of the peak of the continuum and a point about  $\sim$2\,kpc away from the nucleus, inside the jet-like feature. The top spectrum corresponds to the nucleus, while the bottom one is an extranuclear spectrum. The area demarked by the big red square corresponds to the region explored at this work.}
    \label{fig:large}
\end{figure*}

\subsection{Stellar Populations}

The optical spectra of powerful AGN host are typically dominated by two main components: a Featureless Continuum (FC), where the spectral energy distribution is reproduced by a power-law $F_{\lambda} \propto \nu^{-\alpha}$; and a continuum emission yielded by the Stellar Population (SP). To further explore this components, it's necessary to decompose each spectrum into the aforementioned contributions. For this purpose, a Stellar Population Synthesis (SPS) code can be used.  Besides enabling the study of stellar populations, the subtraction of the stellar population contribution from the observed spectra is crucial to accurately account for the properties of emission line gas. This is particularly important for emission lines such as H$\alpha$ and H$\beta$, which can be significantly absorbed if a young stellar population dominates the observed continuum emission.

For the stellar population fits we employed the \textsc{starlight} code \citet{2005MNRAS.358..363C}, \citet{asari2007}, \citet{cid2018hahaha}. The code can be easily customized to the MUSE data. This code uses a linear combination of simple stellar populations (SSPs) libraries to fit the continuum of the galactic spectra. It fits the stellar Line Of Sight (LOS) velocity displacement ($v_{*}$) and dispersion ($\sigma_{*}$) for each population fraction ($j$)  by adopting a Gaussian line-of-sight velocity distribution. Additionally, the code provides the option to choose a reddening law to correct for stellar reddening ($A_{\lambda,j}$). Upon completion of the run, the code provides the final model ($M_{\lambda}$) and the contribution of each component ($j$) to the observed flux ($L$). Basically, the $M_{\lambda}$ is given by
\begin{equation}
     M_{\lambda}=\sum_{j=1}^{n}L_{\lambda,j}=\sum_{j=1}^{n}L_{\lambda,j}^{0}\otimes G(v_{*},\sigma_{*})10^{-0.4A_{\lambda,j}}.
     \label{Eq1}
\end{equation}


An important step to perform SPS is the choice of a spectral library.
Given that NGC 232 is a spiral galaxy, we used a customized blend of the \textsc{miles} \citet{2010MNRAS.404.1639V} and \citet{2005MNRAS.357..945G} models. The so called "GM" library (hereafter as GM), is described in \citet{2013A&A...557A..86C} and was updated by \citet{2021MNRAS.501.4064R} and \citet{2023MNRAS.524.5640R} using the most up to date version of \textsc{miles} \citet{2016MNRAS.463.3409V}. This modified library consists of 85 spectra spread in four metallicities: 0.19, 0.40, 1.00 and 1.66 $Z_{\odot}$; and  21 ages: 0.001, 0.006, 0.010, 0.014, 0.020, 0.032, 0.056, 0.1, 0.2, 0.316, 0.398, 0.501 0.631, 0.708, 0.794, 0.891, 1.0, 2.0, 5.01, 8.91 and 12.6 Gyr.
  The spectral resolution of both libraries has been homogenized to match with that of the MILES library \citep[see ][]{2013A&A...557A..86C,CidFernandes14}. This enables precise fits of the spectral features in NGC\,232. Finally, since NGC\,232 is a LIRG, we adopt the extinction law of \citet{1994ApJ...429..582C}.

\subsection{Emission-line profile fitting}

The optical spectra of NGC\,232 present prominent emission lines, as can be seen in Figure~\ref{fig:large}.  In order to measure the gas properties, we fit the emission line profiles by Gaussian curves, after subtracting the contribution of the stellar populations, as derived by the spectral synthesis. We use the Python package \textsc{ifscube} \citep{2020zndo...4065550R,2021MNRAS.507...74R} to  perform the emission-line fitting. By visual inspection of the spectra, we see that asymmetric profiles are detected in several locations, as for instance, blue wings  clearly seen in the [O\,{\sc iii}] emission lines, as shown in Fig.~\ref{fig:large}. These profiles can be properly reproduced by two Gaussian components and thus, we fit each emission line profile by up to two Gaussian functions. The following emission lines were fitted: H$\beta$, [O\,{\sc iii}]$\lambda$5007, [O\,{\sc iii}]$\lambda$4959, [O\,{\sc i}]$\lambda$6300, [N\,{\sc ii}]$\lambda$6548, H$\alpha$, [N\,{\sc ii}]$\lambda$6584, [S\,{\sc ii}]$\lambda$6716 and [S\,{\sc ii}]$\lambda$6731. The blueshifted component in the line profile are physically associated to outflowing gas, while the second component is produced by gas in the plane of the disc.

To reduce the number of free parameters and avoid spurious fits, the subsequent constraints were used: (i) we define two kinematic groups for the Gaussian component representing the disc emission, by keeping tied the centroid velocity and velocity dispersion of the emission lines in the blue side of the spectra (H$\beta$, [O\,{\sc iii}]$\lambda$5007 and [O\,{\sc iii}]$\lambda$4959) and in the red part of the spectra ([O\,{\sc i}]$\lambda$6300, [N\,{\sc ii}]$\lambda$6548, H$\alpha$, [N\,{\sc ii}]$\lambda$6584, [S\,{\sc ii}]$\lambda$6716 e [S\,{\sc ii}]$\lambda$6731), separately. (ii) The blueshifted components were set to have a higher velocity dispersion than the narrow (disc) ones. By using the {\it refit} parameter of the \textsc{ifscube} code, this usually implies that if only the disc component is enough to reproduce the line profile, the amplitude of the second Gaussian component will be set as zero. (iii) the broad components associated to the same element were kept in the same kinematic group (same velocity and width).  The {\sc ifscube} code outputs a multi-extension "fits" file containing the best-fit parameters, fluxes,  equivalent widths (EWs) at each spaxel of the datacube, among others. These measurements are used to construct two-dimensional maps, which are presented in the next section, along with the results for the stellar populations.

\section{Results}\label{sec:results}

\subsection{Stellar Populations}

\begin{figure*}
    \centering
    \includegraphics[width=.85\textwidth]{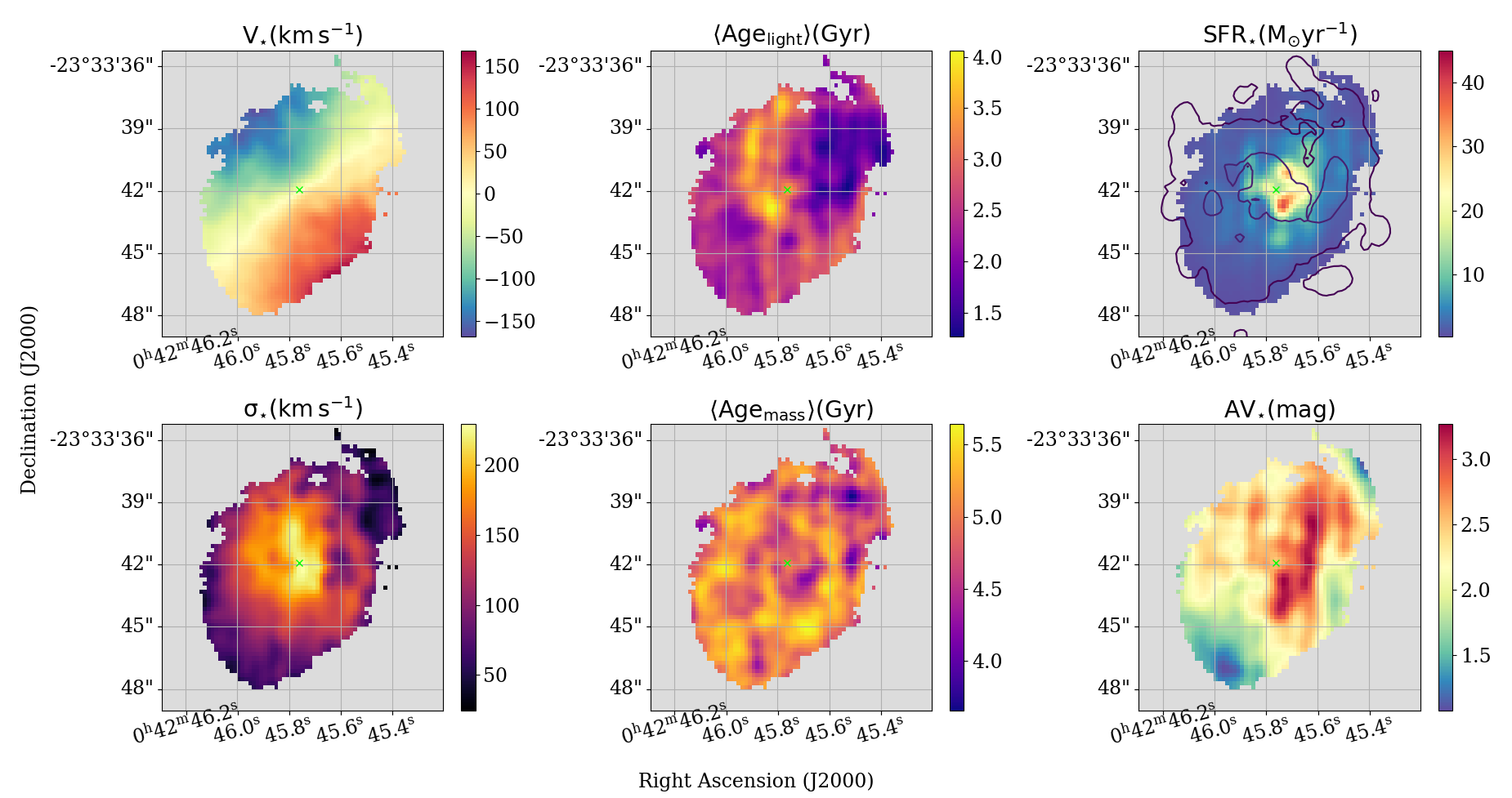}
    \includegraphics[width=.85\textwidth]{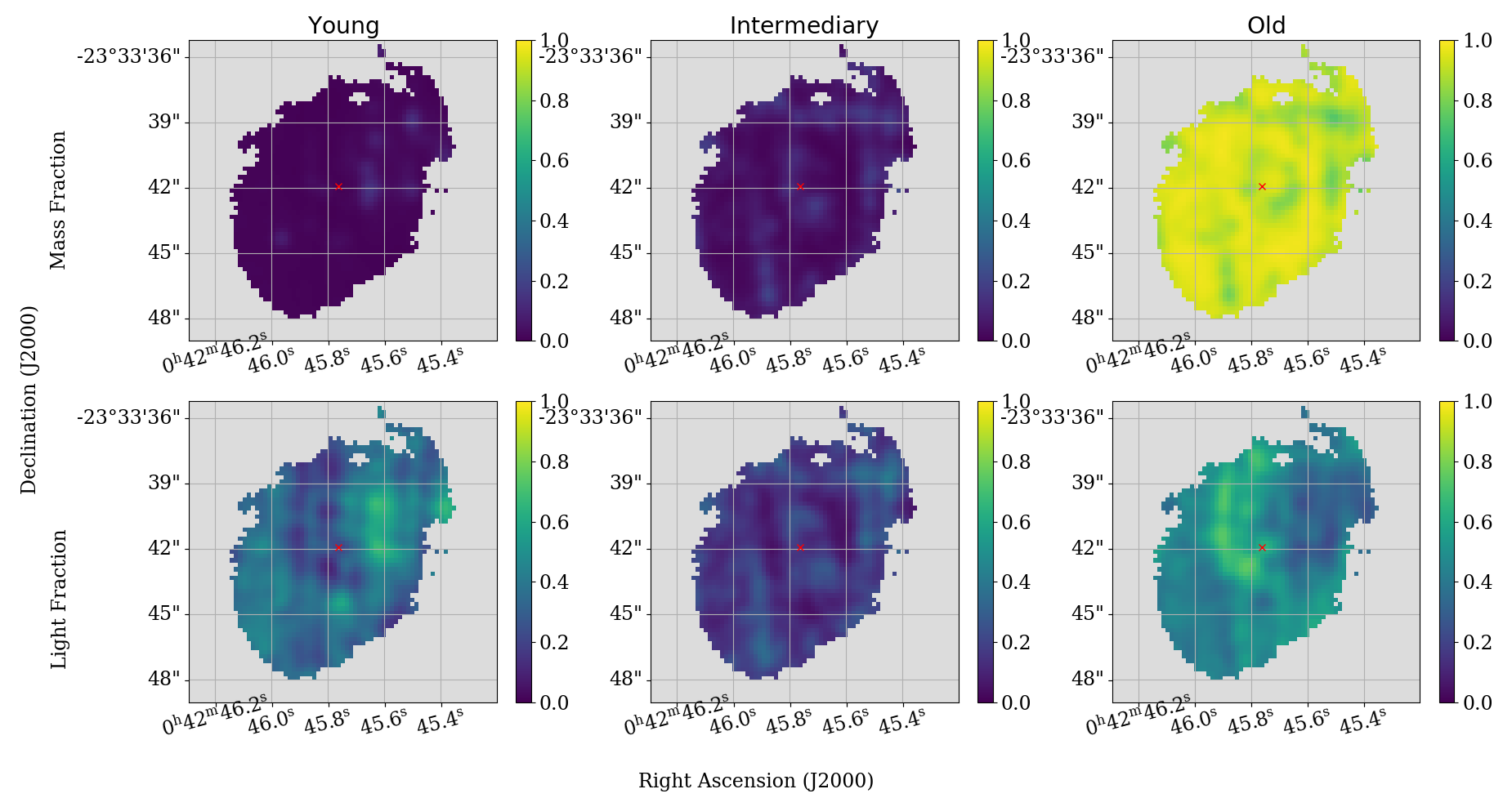}
    
    \caption{Stellar Population Synthesis results. The top panels show, from left to right, the LOS stellar velocity field, the mean age of the stellar populations weighted by the continuum emission, and the stellar SFR derived considering the contributions of the stellar populations younger than 20\,Myr.  H$\alpha$ flux contours are traced in purple as reference. The second row shows, from left to right, the stellar velocity dispersion, mean age of the stellar populations weighted by the mass fraction,  and the stellar visual extinction ($A_{\rm V}$) extinction. The third and bottom rows show the population fraction to each bin. Respectively the mass fraction that dominates each bin (from left to right) young, intermediary and Old. The bottom line, display the same, however for the light fraction. The central cross marks the position of the continuum peak, while the gray regions correspond to the spaxels with $S/N$\,<\,10.  }
    \label{fig:3}
\end{figure*}

In Fig.~\ref{fig:3} we present the resulting maps for the stellar populations mean properties, as well as the population fractions within three age bins: Young ($t\leq$ 50\,Myr); Intermediary, (50\,Myr $< t \leq$ 2\,Gyr) and  Old ($t>$2\,Gyr). We exclude regions where the signal-to-noise (S/N) ratio at the continuum is smaller than 10, a minimum value required to obtain reliable results from the stellar population synthesis \citep{2005MNRAS.358..363C} and the central point indicates the position of the peak of the continuum emission.

The top-left panel shows the stellar velocity field, after subtracting the systemic velocity of the galaxy of 6799 km\,s$^{-1}$, as obtained from the mean value of the stellar velocities within 0\farcs4 radius centred at the peak of the continuum emission. This map shows a clearly rotation pattern with line of nodes oriented approximately along the position angle, PA$\approx30^\circ$, showing blueshifts to the northeast and redshifts to the southwest of the nucleus, and a projected velocity amplitude of $\sim$150 km\,s$^{-1}$. The stellar velocity dispersion ($\sigma_*$) map, shown in the left panel of the second row of Fig. \ref{fig:3}, presents values in the range from $\sim50$ to $\sim$220 km\,s$^{-1}$, with the smallest values observed to the northwest and to the southeast. The highest $\sigma_*$ values are observed slightly displaced from position of the continuum peak. 

The central panels of first and second rows of Fig.~\ref{fig:3} show the mean age of the stellar populations, weighted by their emission at normalization wavelength of 5870\AA\,(top) and by their mass (second row) contributions. Stellar populations older than $\sim$4.5 Gyr dominates the mass contribution at most locations, with a mass-weighted mean ages in the range $\sim$3.8--5.5\,Gyr. The light-weighted map shows mean ages of $\sim$3.5 Gyr along the galaxies major axis, while mean ages as lower as 1.5 Gyr are observed in regions northwest and southeast of the nucleus, along the galaxy's minor axis.

The map for the star formation rate derived by the spectral synthesis (SFR$_{\star}$) is shown in the top-right panel of Fig.~\ref{fig:3}. Following \citet{2021MNRAS.501.4064R}, we derive the SFR$_{\star}$ over a time $\Delta t$ by 
\begin{equation}
    {\rm SFR_\star} = \frac{\sum_{j_i}^{j_f} M_{\star,j}^{\rm ini}}{\Delta t}.
\end{equation}
To derive ${\rm SFR_\star}$ we consider ${\Delta t}=20\,Myr$, which represent the stellar populations that dominate emission of ionizing photons, presenting also the best correlation with the gas SFR \citet{2021MNRAS.501.4064R}. Such young stellar populations are found only in a few locations to the east-southeast and west of the nucleus. The derived star formation rate are quite high, with typical values of $3-5$ M$_\odot$\,yr$^{-1}$, reaching values of up to 20 M$_\odot$\,yr$^{-1}$ to the southeast of the nucleus. For comparison, we have overlaied contours from the H$\alpha$ narrow component flux distribution, showing that most of the regions with contributions of young stellar populations are associated to structures seen in the H$\alpha$ flux map.

Another result from the synthesis is the visual extinction $A_{V\star}$ map (second row, right panel of Fig.~\ref{fig:3}), which traces the dust content where stellar populations are embedded. The $A_{V\star}$ map shows values ranging from 1.3 to 3.0 mag, with the lowest values observed to the southeast of the nucleus and highest ones to the northwest. This indicates that the near side of the disc corresponds to the northwestern side of the galaxy. In addition, some high obscuration is also observed in a structure with $\sim 3\arcsec$ extension from the nucleus, observed to the south, co-spatial with the structure of highest star-formation rate, seen in the top-right panel.

The third and bottom rows of Fig.~\ref{fig:3} show the maps for the contribution of stellar populations within three age bins, labeled as: Young ($t\leq$ 50\,Myr); Intermediary, (50\,Myr $< t \leq$ 2\,Gyr) and  Old ($t>$2\,Gyr). This binning is done in order to avoid any age-metallicity degeneracy of the SSPs, that may be occasioned by the adjacent noise in the data, as suggested by \citet{2004MNRAS.355..273C, 2009MNRAS.400..273R, 2021MNRAS.501.4064R,2022MNRAS.512.3906R,2023MNRAS.524.5640R}. The mass-weighted contribution is dominated by the old stellar populations, while in the emission-weighted maps, the highest contributions are seen for the young and old stellar populations. 


\subsection{Gas emission distributions and kinematics}

\begin{figure*}
    \centering
    \includegraphics[width=.98\textwidth]{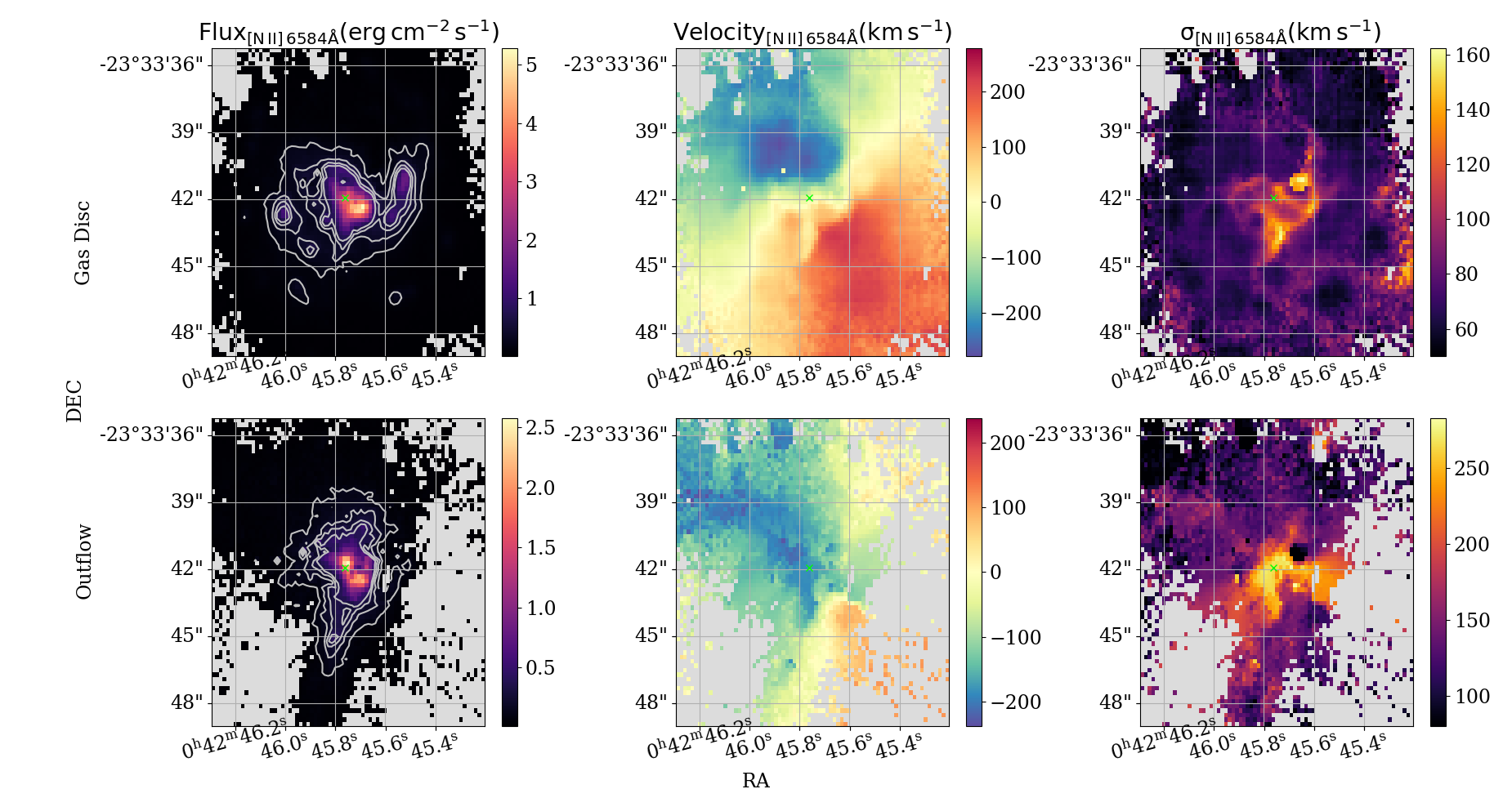}
    \includegraphics[width=.95\textwidth]{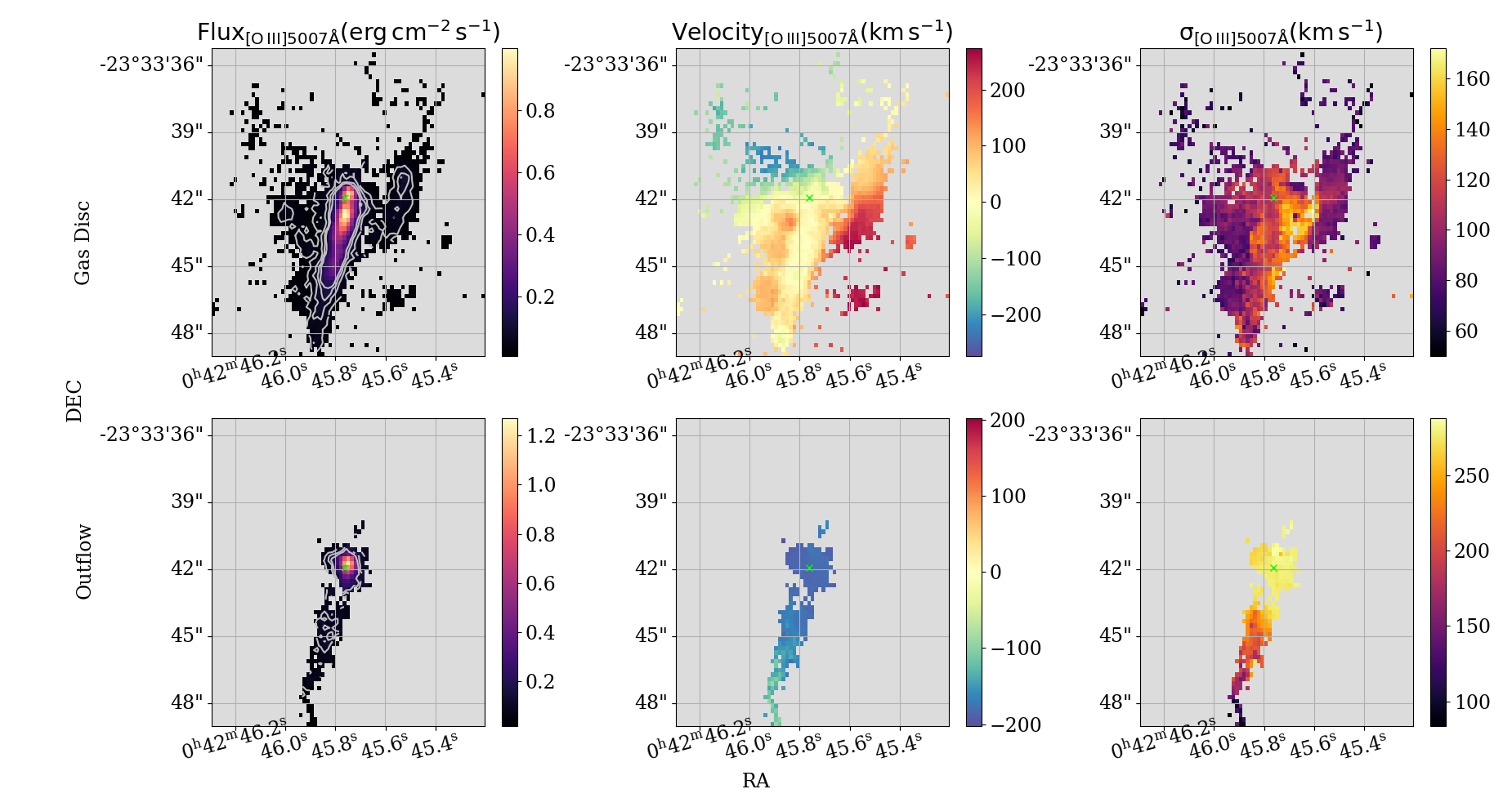}
    \caption{At the right column, the integrated fluxes for the lines of [N\textsc{ii}]$\,\lambda 6584$ and [O\textsc{iii}]$\,\lambda 5007$ are shown. Both disc and outflow components, are indicated at the right side of the figure. The white contours correspond to flux levels for each line, aiming on highlight fainter structures. The color bars show the fluxes in units of $10^{-16}\,erg\,s^{-1}\,cm^{-2}$. The central column, displays the LOS centroid velocity fields for each component. The color bars show the velocities, in km\,s$^{-1}$, relative the systemic velocity of the galaxy. The right panels show the velocity dispersion maps for each line and component, with color bar showing the values in km\,s$^{-1}$. The gray areas, correspond to points where the line amplitude is bellow 3-$\sigma$, and the central cross, tags the center of emission.   }
    \label{fig:gas}
\end{figure*}

As mentioned above, we fit the emission-line profiles by two Gaussian functions: a narrow component, originated mostly from emission of gas in the galaxy disc, and a broad component, tracing an outflow. Fig.~\ref{fig:gas} presents the flux distributions and kinematic maps for the [N\,{\sc ii}]$\lambda6584$ and [O\,{\sc iii}]$\lambda5007$ emission lines. The left column shows the flux distributions, the velocity fields are shown in the central panels and the velocity dispersion maps are presented in the right panels. From the top to the bottom, results for the narrow [N\,{\sc ii}]$\lambda6584$, broad [N\,{\sc ii}]$\lambda6584$, narrow [O\,{\sc iii}]$\lambda5007$ and broad [O\,{\sc iii}]$\lambda5007$ components are shown, respectively. In all panels, white regions represent masked spaxels where the amplitude of the corresponding emission line is lower than three times the standard deviation in the continuum calculated in a region close to the emission line. We show the flux maps for the other emission lines in Fig.~\ref{fig:ap_flux}. 

The flux distribution of the [N\,{\sc ii}]$\lambda6584$  narrow component is similar to those observed for H$\alpha$, [O\,{\sc i}]$\lambda6300$ and [S\,{\sc ii}]$\lambda\lambda6716,6731$ emission lines.  The peak of the emission line fluxes is observed $\approx$1\arcsec southwest of the position of the continuum peak (indicated by the cross) and patches of enhanced emission, following the spiral structure seen in inner $\sim$3$^{\prime\prime}$. The higher ionization gas emission, traced by the [O\,{\sc iii}]$\lambda5007$ line, is mainly observed along a jet-like structure towards the southeast, as previously reported by  \citet{2017ApJ...850L..17L}. In addition, the narrow [O\,{\sc iii}]$\lambda5007$ shows some emission to the west, in a region co-spatial with emission structure seen in the flux map for the [N\,{\sc ii}]$\lambda6584$ narrow component. The broad  [N\,{\sc ii}]$\lambda6584$ flux distribution, besides showing emission along the jet-like structure, presents extended emission to the northeast. The morphology of the  [N\,{\sc ii}]$\lambda6584$ (and other low-ionization emission lines) narrow component can be described as a ``V-shaped" structure, most extended to the east side of the nucleus. Some extended emission in the  [N\,{\sc ii}]$\lambda6584$ broad component is also observed to the northwest of the nucleus.

The velocity field for the narrow component shows a similar rotation pattern as observed for the stars (Fig.~\ref{fig:3}), showing blueshifts to the northeast and redshifts to the southwest of the nucleus. The projected gas velocity amplitude of $\sim$250\,km\,s$^{-1}$, is about 100\,km\,s$^{-1}$ higher than that observed for the stars. This is expected due to projection effects, once the gas is located in a thinner disc than the stars. The [O\,{\sc iii}] broad component velocity in the jet-like structure is seen blueshifted by $\sim$200~km\,s$^{-1}$, while velocities close to zero are seen for the  [N\,{\sc ii}]$\lambda6584$. The extended emission from the  [N\,{\sc ii}]$\lambda6584$  broad component to the northeast is seen blueshifted relative to the systemic velocity of the galaxy. To the southwest and northwest, the [N\,{\sc ii}]$\lambda6584$  broad component velocity map show values close to zero, slightly redshifted at some locations. 

The velocity dispersion ($\sigma$) maps for the narrow component show values ranging from $\sim60$ to 160 km\,s$^{-1}$. The overall small $\sigma$ values is consistent with emission of gas in the plane of the disc. In the inner $\sim$3$^{\prime\prime}$ radius, the highest $\sigma$ values delineate a ``V-shaped" (or conical) structure oriented along the minor axis of the galaxy. The broad component show $\sigma$ values larger than 150\,km\,s$^{-1}$ in most regions, with the highest ones seen in the location of the jet-like structure. This indicates that the broad component is tracing a more turbulent, outflowing gas.

\subsection{Emission-line ratio diagnostic diagram}

We can use the optical emission line ratio diagnostic diagrams to investigate the excitation origin of the gas emission in NGC\,232. The most widely used diagnostic diagram is the [N\,{\sc ii}]-based BPT diagram, which involves the [O\,{\sc iii}]$\lambda5007$/H$\beta$ versus [N\,{\sc ii}]$\lambda6584$/H$\alpha$ flux ratios \citep{bpt}. In Figure~\ref{fig:bpt}, we present the  [N\,{\sc ii}]-based BPT diagram for the narrow (top-left panel) and broad (bottom-left panel), along with their excitation maps, spatially identifying different regions of the diagram. { The separating line proposed by \citet{kewley01} is based on photoionization models and demarcates the upper limit of flux line ratios that can be reproduced by ionizing photons from young stars, while the line of \citet{kauffmann_2003} is based on the observed distributions obtained from the SDSS integrated spectra. However,  given that the majority of AGNs also harbor star-forming regions, emission line fluxes based on integrated spectra of AGN hosts may include contamination from these star-forming regions. Consequently, the object's position on the BPT diagram may be influenced by the varying mixture of contributions from both AGN and star formation to the ionization of the gas \citep{Dagostino19c,Agostino21}. This is a problem that can be alleviated with spatially resolved observations \citep[e.g.][]{Dagostino19a,Dagostino19b,Ji20,Ji23,Law21,Agostino23}. In Fig.~\ref{fig:bpt} we present the 
 empirical boundaries from \citet{Law21}, established through the analysis correlations between the gas velocity dispersion and flux line ratios measured for 3.6 million spaxels distributed across 7400 individual galaxies observed as part of the MaNGA survey. This diagram is particularly well-suited for spatial excitation maps because its development depend on "kinematic temperature" rather than photoionization models integrated emission.  The excitation maps shown in the right panels of Fig.~\ref{fig:bpt} are based on the separating lines proposed by \citet{Law21}. For comparison, we also added the classical limits to integrated data, the dotted line represents the empirical maximum starburst line proposed by \citet{kauffmann_2003}, with values bellow this line being consistent star-forming regions. Points above the dashed line, from \citet{kewley01}, are consistent with AGN ionization, objects located between these two lines are usually refereed as transition objects (TOs). The solid line, from \citet{Cid_Fernandes_2010}, separates weak (e.g. LINER) and strong (e.g. Seyfert) AGN ionization.  In Fig.~\ref{fig:bpt}, we exclude all spaxels where at least one of the lines is not detected above 3-$\sigma$ of the noise level.

The BPT diagram for the disc component show points in all emission types. The flux-line ratios in the jet-like structure are consistent with AGN ionization, with most points located in the region occupied by Seyfert galaxies, surrounded by a LIER-type excitation  \citep{2016MNRAS.461.3111B}, mostly due to the attenuation of the radiation field. Star-formation ionization is observed in regions away from the nucleus, to the west and to the east of it. In regions between the star-forming regions and the jet-like structure, line ratios typical of TO are observed. The BPT for the outflow component is restricted to the region of the jet-like structure, with most of the points located above the line of \citet{kewley01}.

\begin{figure*}
\centering
\includegraphics[scale=.50]{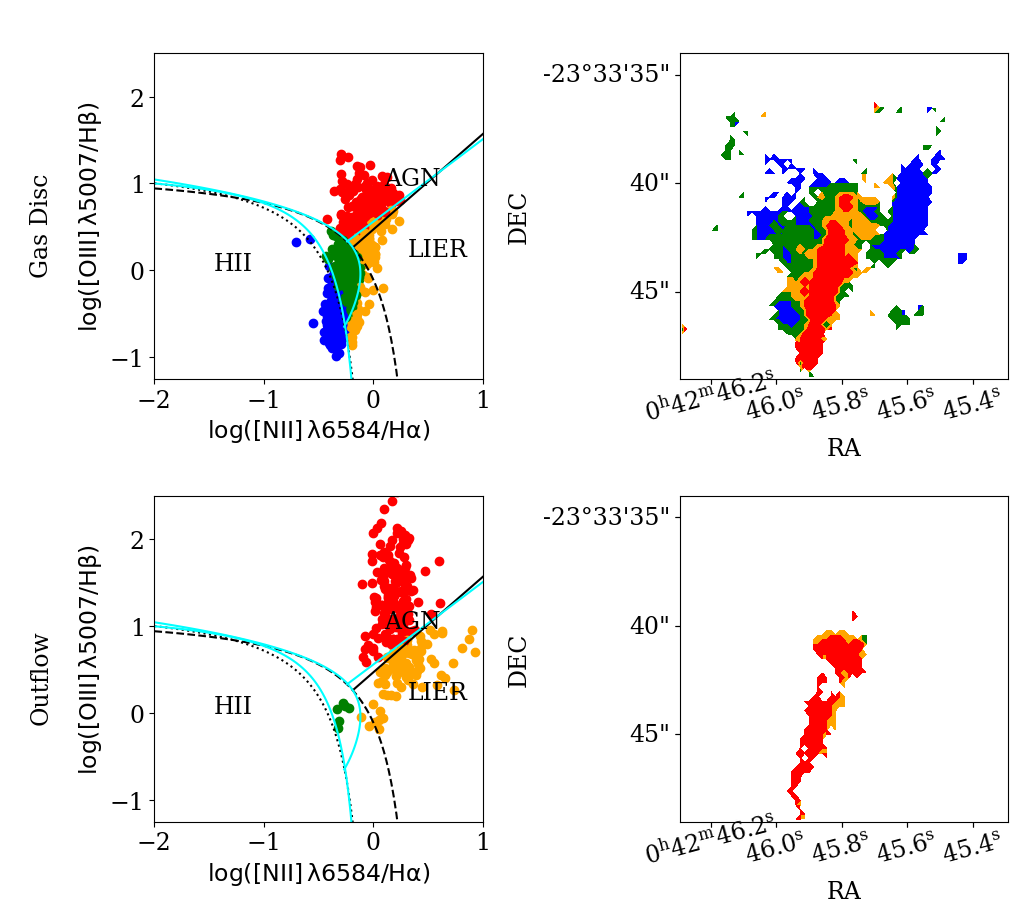}
\caption{[N\,{\sc ii}]-based BPT diagram (left) and excitation map (right) for the disc (top) and outflow (bottom) components. The dotted line corresponds to the empirical limit for star forming galaxies derived by \citet{kauffmann_2003}, while the dashed line is from \citet{kewley01} and sets the upper values that still can be reproduced by stellar ionization \citet{kewley01}. Typical values for AGN are above this line. The solid line is the separation of Strong AGN and Low Ionization Emission-line Region (LIER's) \citep{Cid_Fernandes_2010}. The cyan line delineates the empirical boundaries for identifying the source of ionization within individual spaxels, based on kinematic activity \citep{Law21}. }
\label{fig:bpt}
\end{figure*}

\subsection{Gas extinction, density and oxygen abundance}

\begin{figure*}
    \centering
    \includegraphics[width=.33\textwidth]{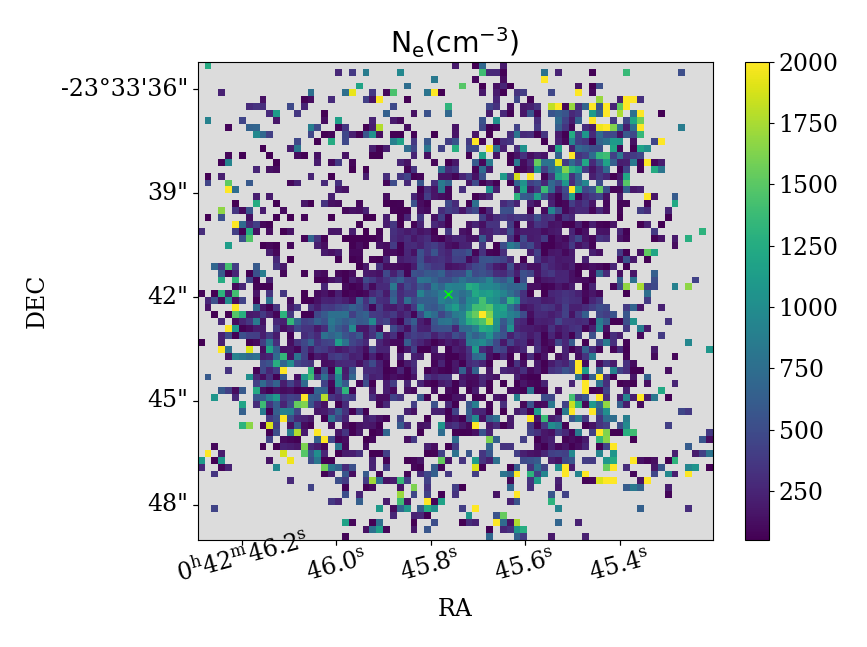}
    \includegraphics[width=.33\textwidth]{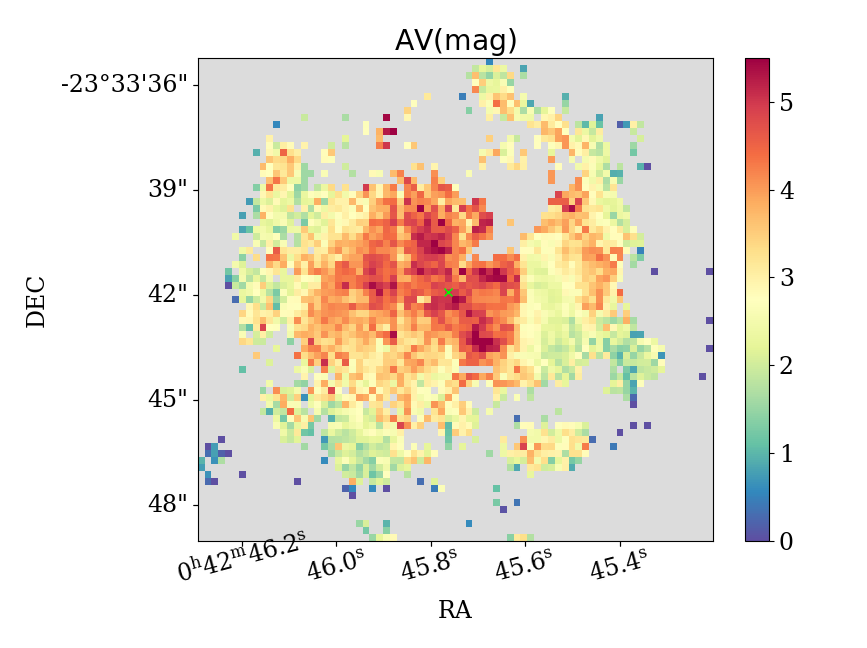}
    \caption{Electron density map (left) obtained from the [S\,{\sc ii}] doublet. The central panel shows the visual extinction ($A_{\rm V}$ map obtained from the H$\alpha$/H$\beta$ line ratio.}
    \label{fig:ratio}
    
\end{figure*}

We estimate the electron density ($N_{\rm e}$) using the [S\,{\sc ii}]$\lambda$6716/[S\,{\sc ii}]$\lambda$6731 observed line ratios. The broad component of the [S\,{\sc ii}] lines is detected, with S/N > 3, only in a few locations, despite the presence of asymmetric wings in the line profiles. Thus, to estimate the $N_{\rm e}$, we use the total [S\,{\sc ii}] fluxes at each spaxel. To reduce the number of free parameters in the fitting of the line profiles, we reproduce the lines with Gauss-Hermite series, which properly account for deviations of the observed profiles relative to a Gaussian curve. We use the \textsc{pyneb} Python package \citet{2015A&A...573A..42L} and adopt an electron temperature of 15\,000\,K, a typical value for AGN hosts \citep{revalski18a,revalski18b,dors20,rogemar21_te,rogemar21_tflut,2023MNRAS.520.1687A}.

The electron density map is displayed in the left panel of Figure \ref{fig:ratio}. The gray regions represent masked locations, where at least one of the [S\,{\sc ii}] lines is not detected above 3-$\sigma$ noise level of the continuum, as well as the values out of the method boundaries \citep[i.e. ${N_{\rm e}}\sim50\,{\rm cm^{-3}}$ and ${N_{\rm e}}\sim2.7\times10^4,{\rm cm^{-3}}$][]{2006agna.book.....O}. The $N_{\rm e}$ values are in the range from $50$ to 2000\,cm$^{-3}$, with the highest values observed close to the nucleus, slightly displaced towards the southwest.  The mean value value is $\langle {N_{\rm e}} \rangle =747\pm254\, {\rm cm^{-3}}$.

The H$\alpha$/H$\beta$ flux line ratio can be used to calculate the gas extinction. The visual extinction parameter can be obtained by
\begin{equation}
    A_{V}=7.22\,\log \left(\frac{F_{\rm H  \alpha}/F_{\rm H  \beta}}{2.86} \right),
\end{equation}
where $F_{\rm H\alpha}$, $F_{\rm H\beta}$ are the observed fluxes and we have adopted a theoretical ratio between H$\rm \alpha$ and H$\rm \beta$ fluxes of 2.86, corresponding to the case B recombination at the low density regime ($N_{\rm e}\,=\, 100$) and an electron temperature of 10\,000 K \citep{2006agna.book.....O}. As for the electron density estimation, we use the total line fluxes. The resulting $A_{\rm V}$ map is shown in the middle panel of Fig.~\ref{fig:ratio}. Although the H$\rm \alpha$ emission is detected over most of the field of view, the $\rm H\beta$ is not detected in regions closer to or further away from the nucleus. These locations are masked out and shown in gray. The $A_{\rm V}$ map shows values ranging from 0 to 4.5 mag, with the highest values observed in the inner 3$^{\prime\prime}$ and the lowest at larger distances from the nucleus. 

Finally, we can also use the observed emission-line flux ratios to estimate the oxygen abundance, using the strong line method, hence auroral lines (e.g. [O \textsc{iii}]$\lambda$ 4363) are out of the spectral coverage, making unable to apply the $T_{\rm e}$-method. There are available calibrations for star forming regions and for AGN, and thus, our analysis is restricted to spaxels located in these two regions of the classical BPT diagram.  For the AGN like emission, we used the calibration from \cite{1998AJ....115..909S} obtained by using AGN photoionization models to reproduce observed optical emission lines from Seyfert nuclei, given by
\begin{equation}
\begin{aligned}
    12\,+\,\log({\rm O/H}) = &8.34+(0.212\,x)-(0.012x^{2})-(0.002\,y)+\\
                    &+(0.007\,xy)-(0.002x^{2}y)+(6.52\times10^{-4}y^{2})+\\
                    &+(2.27\times10^{-4}xy^{2})+(8.87\times10^{-5}\,x^{2}y^{2}),
\end{aligned}                   
\end{equation}
where $x\,=\,[N\textsc{ii}]\lambda\lambda\,6548,6584/H\alpha$, and $y\,=\,[\rm O\textsc{iii}]\lambda\lambda\,4959,5007/H\beta$. The calibration inside the range $8.4 \leq 12 + \rm log({\rm O/H}) \leq 9.4$, should give results with $\sim$ 0.1 dex inside the confidence interval.

For spaxels in the SF region of the classical BPT diagram, we use the empirical relation of \citet{2009MNRAS.398..949P} that uses the parameter ${\rm O3N2}$ as a calibrator as follows the relation:
\begin{equation}
    12\,+\,\log({\rm O/H})\,=\,8.74\,-\,0.31\,\times\,{\rm O3N2}
\end{equation}
where
\begin{equation}
    {\rm O3N2\,=\,log\left(\frac{[O\textsc{iii}]\lambda\,5007}{H\beta}\times \frac{H\alpha}{[N\textsc{ii}]\lambda\,6584}\right) }.
\end{equation}
 This relation is valid for 12 + $\log({\rm O/H})$ $\gtrsim$ 8.0, since the parameter ${\rm O3N2}$ do not correlate well to low metallicity values. To the "high" metallicity limit, the relation will give results within 0.32 dex\footnote{For a review on discrepancies in O/H abundances derived through distinct methods for AGN see \citet{2020MNRAS.492..468D}, and for SFs see \citet{2010arXiv1004.5251L}.}.

 The resulting O/H abundances present values very similar, with mean values of $\langle 12 + \log({\rm O/H})\rangle=8.80\pm0.05$ for the SF dominated spaxels and  $\langle 12 + \log({\rm O/H})\rangle=8.73\pm0.09$ for the AGN dominated region. The mean oxygen abundance for the star forming regions in the central region of NGC\,232 is similar to values reported for circumnuclear star forming regions in nearby galaxies \citep{2007MNRAS.382..251D,dors08}. Similarly, for AGN dominated spaxels the derived abundances are consistent with O/H estimates for nearby Seyfert galaxies \citep[e.g.][]{dors20}.

\section{Discussion}\label{sec:disc}

In the previous section we have presented two-dimensional maps for the stellar and gas properties. The stellar velocity field shows a well organized rotation pattern (Fig.~\ref{fig:3}), while the gas presents two kinematic component (Fig.~\ref{fig:gas}): (i) a rotating disc, represented by a narrow component in the emission-line profiles, and (ii) an outflow, identified as a broad component in the line profiles. The outflow geometry is consistent with a bi-cone, oriented approximately perpendicular to the major axis of the galaxy, with the northeastern side being in front of the disc and slightly pointing towards us. In this section, we characterize the disc and outflow components, investigate the impact of the outflow in star formation of the host and discuss the star formation history of NGC\,232.

\subsection{The disc component}

We used the \textsc{kinemetry} method \citep{2006MNRAS.366..787K} to model the stellar and gas velocity fields. 
The code uses a surface photometry technique by fitting ellipses along the Line Of Sight Velocity Distribution (LOSVD), finding the orientation of the major axis by using a cosine law. The results of this fit, are presented in Figure \ref{fig:pa}. The model reproduces well the observed velocity field with residual velocities smaller than 10~km\,s$^{-1}$ at most locations. The orientation of the major axis of the fitted ellipses is approximately constant, ranging from $\sim$25$^\circ$ in the inner 0.5 kpc ($1\farcs1$) to $\sim$35$^\circ$ at distances  around $\sim$1.8 kpc ($3\farcs9$)  from the nucleus.  The mean value for the kinematic position angle obtained for the stars is $\Psi_{0}=28^{\circ}\pm4^\circ$. This value is distinct from those obtained for the large scale disc of NGC\,232, e.g., of  $\Psi_{0}=171^{\circ}$ \citep{1989spce.book.....L} using "ESO-LV Quick Blue" IIa-O photographic plates and  $\Psi_{0}=10^{\circ}$ \citep{2003AJ....125..525J} based on Ks-band.
The  $k_5/k_1$ ratio, which measures deviations beyond pure rotation in higher orders, for the stars is small ($k_5/k_1\lesssim0.05$) over all radii. 

The  \textsc{kinemetry} model is able to reproduce the gas velocity field for the disc component over most of the observed field of view, except for regions within the inner $\sim3^{\prime\prime}$ 1.39\,kpc) radius where residuals of up to 60\,km\,s$^{-1}$ are observed. This is also seen in the $k_5/k_1$ ratio radial plot (bottom-right panel of Fig.~\ref{fig:pa}) that shows values larger than those for the stars in the inner region. In addition, the derived kinematic position angle for the gas is misaligned by up to 20$^\circ$ relative to the stellar value (bottom-central panel). The highest residual velocities are observed along the southeast-northwest direction, within the ionization cone and bipolar outflows revealed by the broad line components. We interpret the structures with the largest velocity residuals as being produced by the interaction of outflowing gas with the disc gas. Further support for this interpretation comes from the higher velocity dispersion values observed in the same location for the narrow component (Fig.~\ref{fig:gas}), indicating a more turbulent gas. An alternative interpretation for the significant discrepancies between the observed and modeled gas velocity fields in the inner region could be that the object underwent a recent merger. Misalignment between the gas and star semi-axes is often linked to a sudden inflow of gas, as suggested by \citet{1992ApJ...401L..79B}, \citet{2016MNRAS.457..272D}, and \citet{2021A&A...650A..34R}.  Finally, the bottom-left panel of Fig.~\ref{fig:pa} shows that the gas velocity amplitude is up to 100\,km\,s$^{-1}$ larger than that of the stars. This result is consistent with the fact  that gas is typically situated in a thin disc, whereas the stellar velocity field traces a thicker disc and the stellar bulge, resulting in lower velocities for the latter.

\begin{figure*}
    \includegraphics[scale=.39]{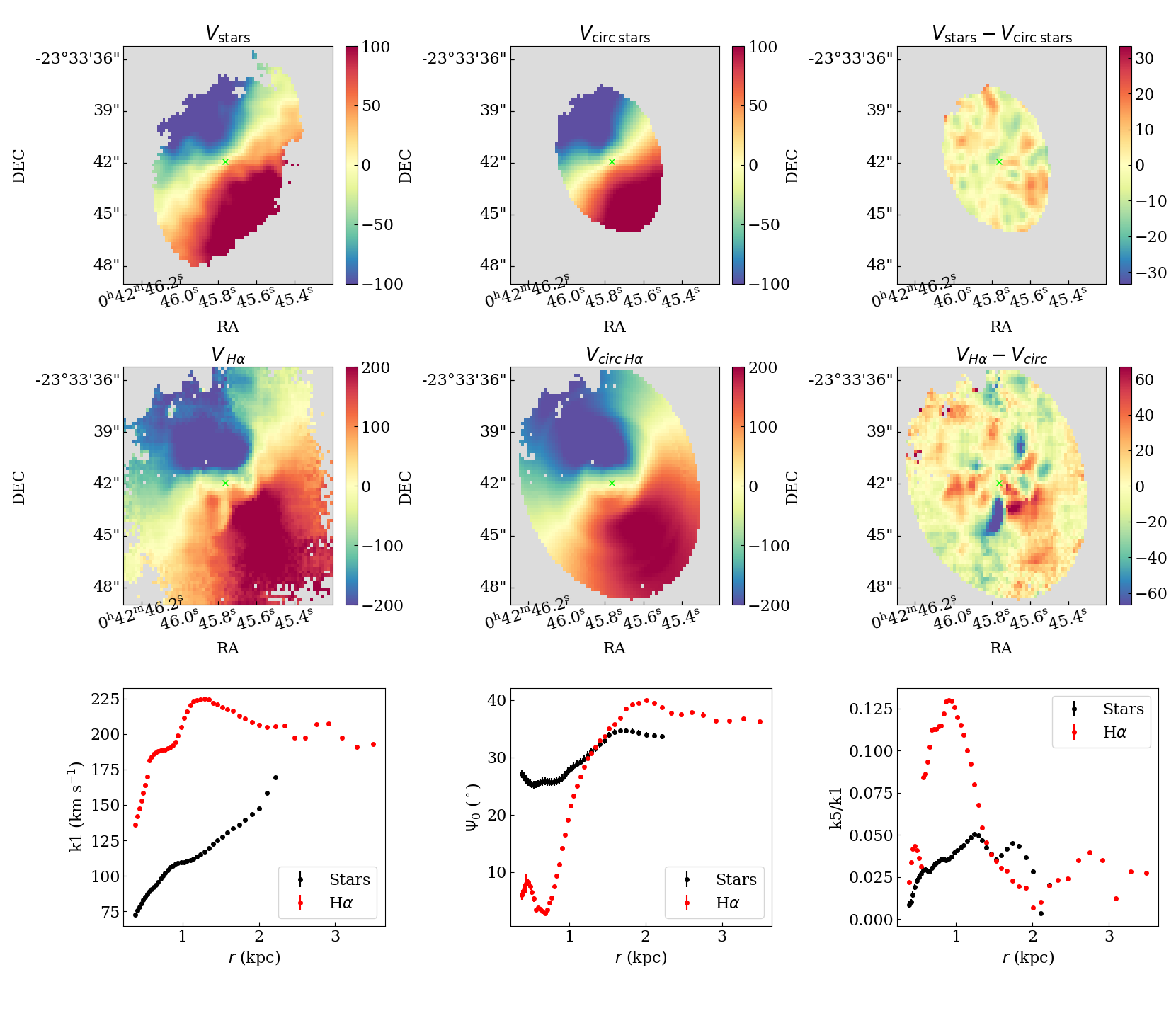}
    \caption{Kinemetry best fit to stars and H$\alpha$ velocity field. The top row shows from left to right: observed stellar velocity field, kinemetry best-fit model, and the residual velocity obtained by subtrancting the model from the observed velocity field. The central row shows the same, but for the gas, traced by the H$\alpha$ narrow component. Gray regions correspond to spaxels masked by the criteria mentioned in the Sec. \ref{sec:results}, and central cross marks the center of emission. The bottom line displays the deprojected velocity for both stellar (in black) and gas (in red) components (left panel), the radial distribution of the position angle of the line of nodes (central panel) and the radial profile of the $k_{5}/k_{1}$ ratio.  }
    \label{fig:pa}
\end{figure*}

\subsection{The outflows}

\begin{figure}
    \centering
    \includegraphics[width=.35\textwidth]{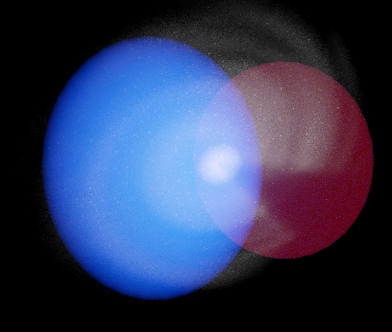}
    \includegraphics[width=.35\textwidth]{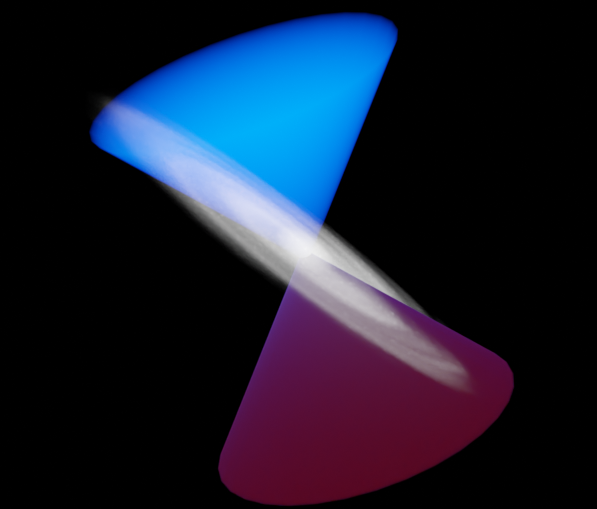}
    \caption{Schematic representation of the disc and outflow in NGC\,232, as seen in our line-of-sight (top panel) and with the disc almost edge-on view (bottom). The disc is inclined by $i=44^\circ$ relative to the plane of the sky, with the major axis oriented approximately along $\Psi_0=28^\circ$. The blueshifted side of the cone is seen in front of the plane of the disc, while the redshifted side is hidden by the disc.} 
    \label{fig:esquema}
\end{figure}

The broad emission-line component is interpreted as being due to an outflow in ionized gas. In high-ionization gas, traced by the [O\,{\sc iii}] emission, this component is detected only in the region co-spatial with the jet-like structure, previously reported by \citet{2017ApJ...850L..17L}. On the other hand, the lower ionization gas, as traced by the [O\,{\sc iii}] emission, exhibits a flux distribution for the broad component that displays a "V-shaped" emission structure, or a cone with its axis oriented towards the southeast. The velocity fields for the broad components show mostly blueshifts relative to the systemic velocity of the galaxy.  Additionally, redshifts are observed towards the northwest and southwest of the nucleus, within areas that appear to correspond to the redshifted counterpart of the cone. Assuming that the spiral arms in NGC\,232 are of the trailing type, from the analysis of the large scale image (Fig.~\ref{fig:large}) and the orientation of the velocity field field (Fig.~\ref{fig:3}), we conclude that the northwest corresponds to the near side of the disc. This interpretation is further supported by the larger extinction values observed at this side in the $A_V$ map of Fig.~\ref{fig:3}.

From the observed gas kinematics and disc orientation, we interpret the broad line component as being due to a biconical outflow, with the southeastern side of the cone pointing towards us, in front of the plane of the disc, and the northwestern side pointing away from us, behind the plane of the disc. The jet-like emission structure, seen in [O\,{\sc iii}], seems to be tracing the emission of cone wall in front of the disc, at high latitudes from the plane of the disc. The northwestern side of the bicone is only partially observed due to dust extinction by the disc.  To calculate the mass-outflow rate in ionized gas, we adopt a biconical geometry as follows. We measure the cone opening angle, of $\sim\,90^{\circ}$, directly from the [N\,{\sc ii}] broad component emission structure (Fig.~\ref{fig:gas}). The disc inclination angle, relative to the plane of the sky, is $i=44^\circ$ obtained by fitting ellipses to the continuum image. The high residual velocities (Fig.~\ref{fig:pa}) in regions within the bicone indicate that the outflowing gas is intercepting the disc. Thus, we assume that the back wall of the southeastern cone and the front wall of the northwestern cone are approximately aligned with the plane of the disc. This implies that the inclination of the bicone axis is $i_{\rm cone}\,\approx\,45^\circ-44^{\circ}=1^{\circ}$ and the front (far) wall of the cone to southeast (northwest) is makes an angle of 46$^\circ$ (-46$^\circ$) relative to the plane of the sky. In Figure~\ref{fig:esquema} we present an schematic representation of the orientation of the disc and bicone in NGC\,232.

 The flux of matter crossing an area, can be found trough:
\begin{equation}
    \dot{M}_{\rm out}=N_{ \rm e}\,v_{\rm out}\, A\, m_{\rm p} f,
\end{equation}
where $N_{\rm e}$ is the electron density, $v_{\rm out}$ is the velocity of the outflow, $A$ is the cross section area, $f$ is a filling factor, and  $m_{\rm p}$ is the proton mass. To eliminate the filling factor, we can use the relation of the $\rm H\alpha$ luminosity ($L_{\rm H\alpha}$), and the volume of the ionized gas

\begin{equation}
    L_{\rm H\alpha}\,\approx\,4\pi\,f\,j_{\rm H\alpha}\,V,
\end{equation}
 what give us
\begin{equation}
     \dot{M}_{\rm out}\,\approx\,\frac{N_{\rm e}v_{\rm out}\,m_{\rm p}\,L_{\rm H\alpha}\,A}{4\pi \,j_{\rm H\alpha}\,V}
\end{equation}
in which  $j_{\rm H\alpha}$ is the $\rm H\alpha$ emissivity, and $V$ is the volume of the region. Adopting the case B recombination for the low-density regime and a temperature of 15\,000\,K, we have $4\,\pi\,j_{\rm H\alpha}$= $2.89\times10^{-25}\,N_{\rm e}^{2}$ \citep{2006agna.book.....O}. We compute the mass-outflow rate within conical rings with height of $\Delta r$ and approximate the ratio between the area of the cross section and the volume of the ring by $A/V\approx3/\Delta r$. Replacing the constants and multiplying the result by a factor 1.4 to account the helium mass and by a factor 2 to account for both sides of the bicone, results  in
\begin{equation}
     \dot{M}_{\rm out}\,\approx\,48.6\, \frac{L_{\rm H\alpha}}{N_{\rm e}}\frac{v_{\rm out}}{\Delta r}
\end{equation}
in cgs units. We compute the velocity of the outflow within each conical ring as $v_{\rm out}=\langle v \rangle / {\rm sin} (46^\circ)$, assuming the the blueshifts observed in the velocity field of the broad component are produced by outflowing gas in the front wall of the cone and, $\langle v \rangle$ is the mean velocity of the broad component within each conical ring. The electron density is assumed as the mean $N_e$ value within each ring using the density map shown in Fig.~\ref{fig:ratio}, while $L_{\rm H\alpha}$ is the integrated H$\alpha$ luminosity for the broad component within each ring, corrected by extinction using the $A_{\rm V}$ values from Fig.~\ref{fig:ratio} and the extinction law from \citet{cardelli_1989}.   In Figure \ref{fig:mout} we present the radial profile of the mass-outflow rate in ionized gas, computed using $\Delta r = $0\farcs4, corresponding to the width of two spaxels. The error bars corresponds to the uncertainties in velocity, flux, and electron density propagated to the mass-outflow rate. We find the maximum outflow rate of 1.26$\,\pm$0.16\,M$_{\odot}$\,yr$^{-1}$ at 1\farcs2 ($\sim$560 pc) from the nucleus.  

Similarly, we can evaluate the kinetic power of the ionized outflows in NGC\,232, by following the equation:

\begin{equation}
\dot{K}\,\approx\,\frac{\dot{M}_{\rm out}}{2}(v^{2}_{\rm out}+3\sigma_{\rm out}^{2}),
\end{equation}
 where $\sigma_{\rm out}$ is the mean velocity dispersion of the H$\alpha$ broad component calculated within each ring. As seen in Fig.~\ref{fig:mout}, the maximum kinetic power of the ionized outflow in NGC\,232 is $3.4\pm0.4\times10^{41}$ erg\,s$^{-1}$.

 The mass rate and kinetic power of the outflows in NGC\,232 can be compared to estimates for other AGN with similar luminosities.  We can estimate the AGN bolometric luminosity ($L_{\rm AGN}$) from the reddening corrected [O\textsc{iii}]$\lambda$5007 luminosity ($L_{[O\textsc{iii}]}$) by  $L_{\rm AGN} = 600\,L_{[O\textsc{iii}]}$ \citep{2009MNRAS.397..135K}. In all locations with detected [O\textsc{iii}] emission, the main excitation source is the AGN, as indicated by the BPT diagrams (Fig.~\ref{fig:bpt}). Thus, we assume that the total [O\textsc{iii}] luminosity is due to the AGN ionization. We integrate the [O\textsc{iii}]$\lambda$5007 extinction corrected fluxes of the broad and narrow components, resulting $ L_{\rm AGN}=(1.7\pm\,0.2)\times10^{44}$ erg\,s$^{-1}$. This value is quite similar to what was obtained by fitting the AGN spectral energy distribution (SED), of  $L_{\rm AGN}= 1.02\pm 0.02\times10^{44}$\,erg\,s$^{-1}$ \citep{2023ApJS..265...37Y}. The mass rate and kinetic power of the ionized outflows derived for NGC\,232 are in the range of values  observed for AGN with similar luminosities, as can be seen from the compilation of outflow properties presented in \citet{rogemar23_nifs}.

The observed kinetic power of the outflow corresponds to $\sim$0.6 per cent of the AGN luminosity in NGC\,232 using the SED based luminosity \citep{2023ApJS..265...37Y}. This value is close to the minimum coupling efficiencies required by most of the simulations for AGN feedback to be effective in suppressing star formation \citep[e.g.][]{dubois_horizon_14,Schaye_eagle_15}. Nonetheless, as extensively debated in the literature \citep[e.g.][]{harrison18}, it is unlikely that all of the energy injected by the AGN gets transformed entirely into outflow kinetic power. Indeed, theoretical results for low luminosity AGN indicate that outflows with relative low kinetic powers can be effective in suppressing star formation in the host galaxy \citep{2023arXiv230300826A}. A rough estimate of the mass of the supermassive black hole (SMBH) in NGC\,232 is $\sim2\times10^8$\,M$_\odot$  obtained from the $M-\sigma$ relation \citep{McConnell11}, using a stellar velocity dispersion of 200\,km\,s$^{-1}$.  For the kinetic power of the outflows and the SMBH mass of NGC\,232, the wind model predictions by \citet{2023arXiv230300826A} indicate that the ionized outflow in NGC\,232 is able to suppress more than 10 percent of the ongoing star formation within the host galaxy, if the wind is sustained by at least 1 Myr.

\begin{figure}
    \centering
    \includegraphics[scale=.3]{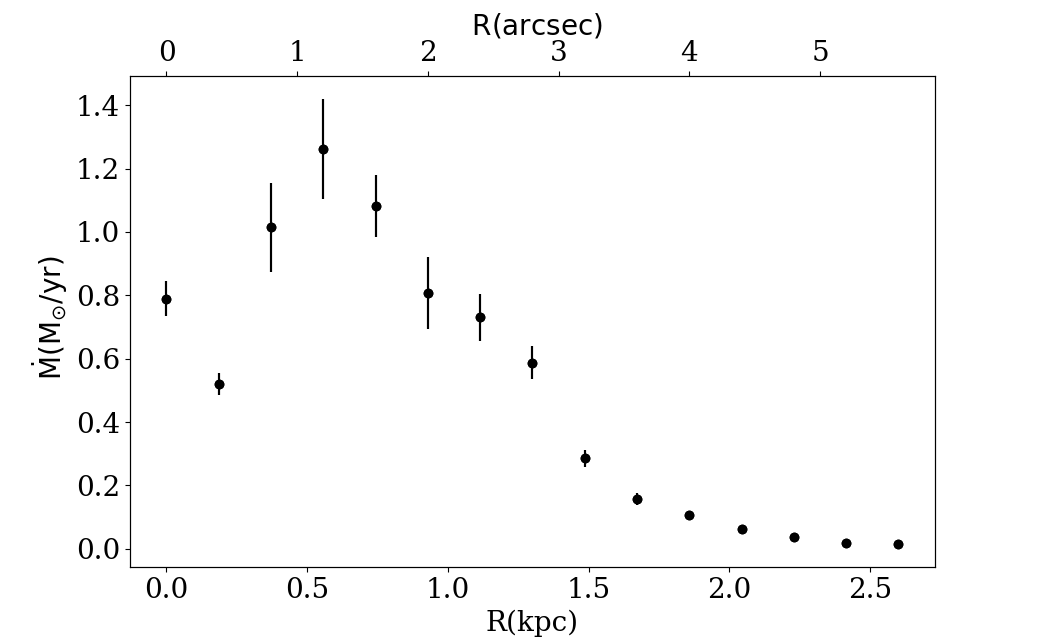}
    \includegraphics[scale=.3]{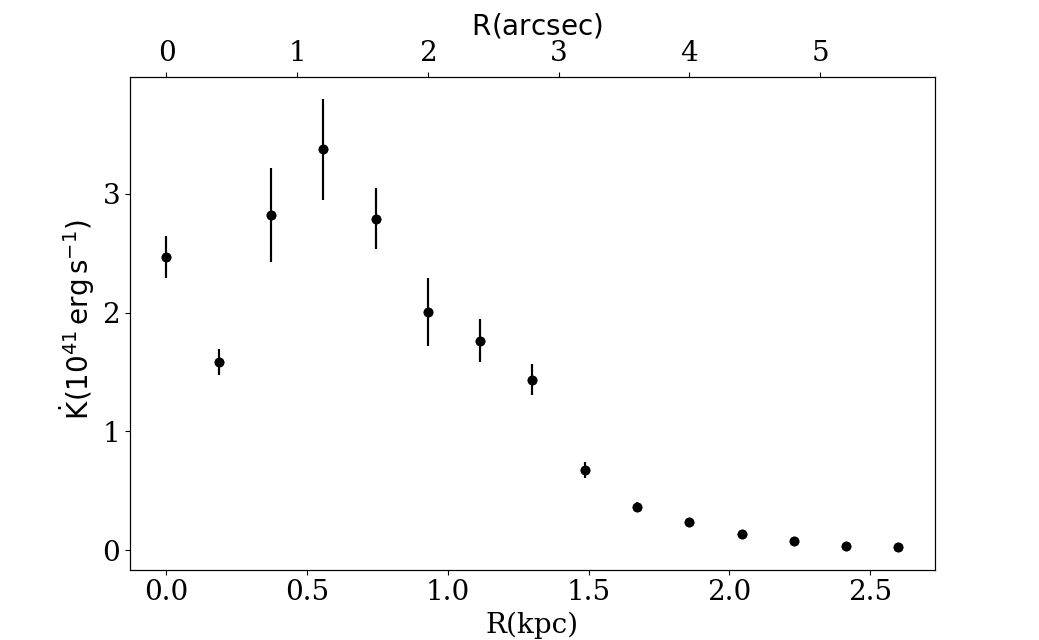}
    \caption{Radial distribution of the mass outflow rate (upper panel) and kinetic power (bottom panel) of the ionized gas outflows in NGC\,232. }
    \label{fig:mout}
\end{figure}

\subsection{Star formation history}

In order to better understand the star formation history in NGC\,232, we constructed radial plots of the contributions of the stellar populations to the observed continuum emission and stellar mass.  We separate the stellar populations in three age bins --   Young ($t\leq$ 50\,Myr); Intermediary, (50\,Myr $< t \leq$ 2\,Gyr) and  Old ($t>$2\,Gyr) -- and calculate their average contributions within circular rings with 0\farcs2 width.  The resulting radial profiles are shown in Figure \ref{fig:lightageradial}. Across all radii, the older stellar populations dominate in both light and mass contributions, with the most significant contribution observed at the nucleus and diminishing as the distance increases. Conversely, the young and intermediary populations exhibit increasing contributions from the nucleus outward. We also present the mean metallicity profile, which displays the lowest values at the nucleus and increases as the distance from it grows.  This is consitent with an inside-out star formation scenario \citep[e.g.][]{perez13,gonzalez-delgado15,diniz17,Bittner20,2018MNRAS.478.5491M}. In addition, the radial increasing gradient of the metallicity, suggests that the central region is mainly dominated by low mass stars, compatible with what was found by \citet{2017MNRAS.466.4731G} for low mass galaxies and for high luminosity AGN, hosted by late type galaxies \citep{2023MNRAS.524.5640R}.  

Finally, the map of star formation rate shown in Fig.\,\ref{fig:3} shows a nuclear ring of SFR$_*$ enhanced values with a radius of $\sim$0\farcs9 (440 pc), similarly to circumnuclear starfoming rings observed in nearby spiral galaxies and more frequently found in AGN hosts than in normal galaxies \citep{hunt_malkan99,hunt99,boker08,dors08,falcao14,rogemar16,hennig17,fazeli20}. The presence of such rings can be interpreted as an evidence of the AGN-starburst connection \citep{Perry85,Terlevich85,norman88,2007ApJ...659L.103R,2019MNRAS.482.4437D} in which both star formation and AGN activity are fueled by the same gas reservoir or, alternatively, the AGN can be triggered by stellar winds from young stars in the ring \citep[e.g.][]{rogemar_09,2007ApJ...670..959S}.  Another potential interpretation for the absence of star formation in the nucleus could be indicative of negative AGN feedback, where the AGN recently quenched the star formation.

\begin{figure*}
\centering
\includegraphics[scale=.26]{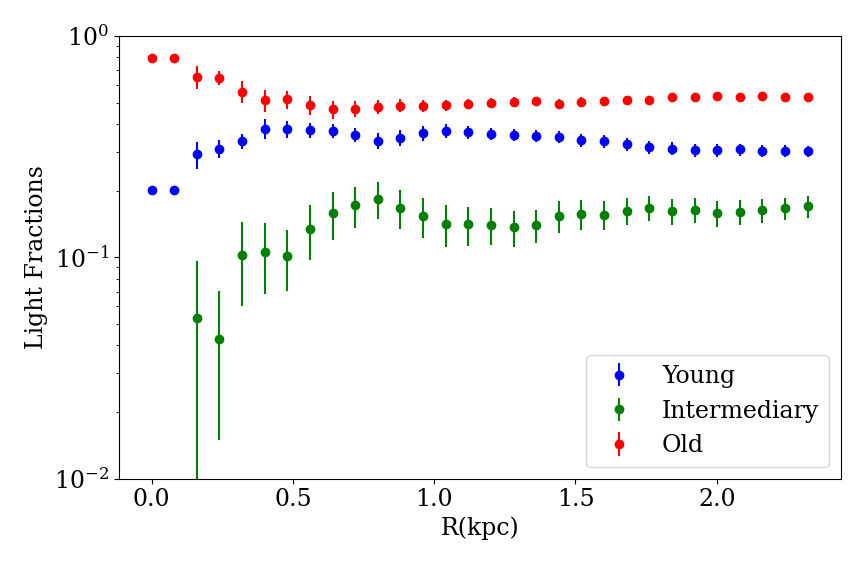}
\includegraphics[scale=.26]{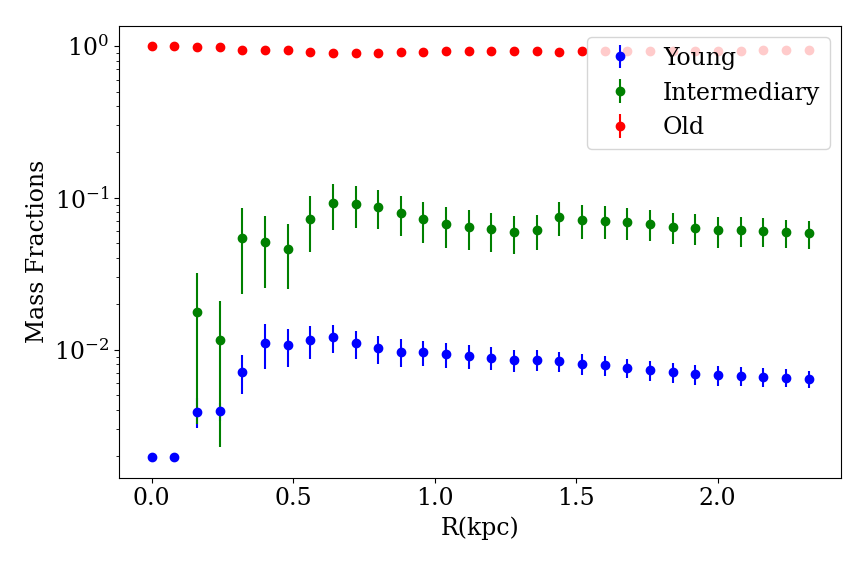}
\includegraphics[scale=.26]{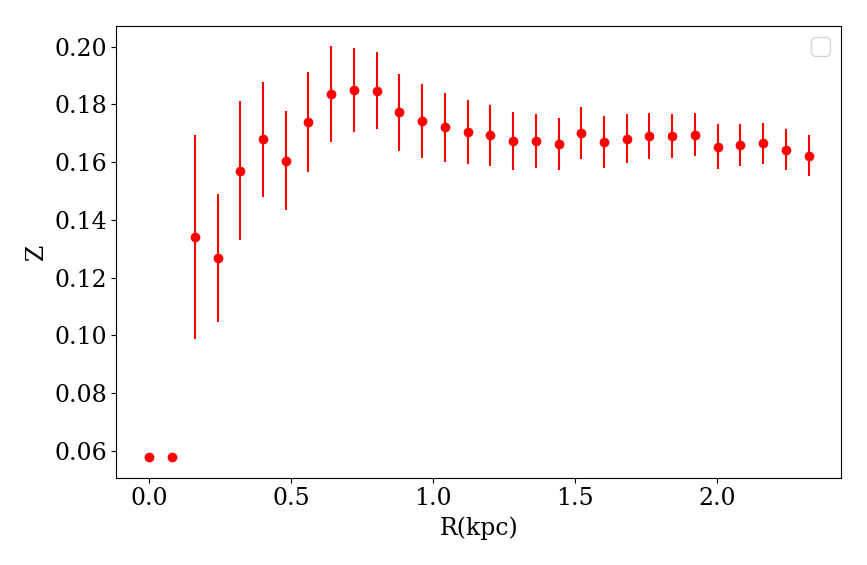}
\caption{Radial distributions of contributions of the young, intermediary and old stellar populations weighted by light  (left panel), and by stellar mass (middle panel). The right panel, shows the radial profile of the stellar metallicity, weighted by mass. }
\label{fig:lightageradial}
\end{figure*}

\section{Conclusions}\label{sec:conc}

We used optical integral field spectroscopy obtained with the Multi Unit Spectroscopic Explorer instrument at the ESO Very Large Telescope to study the stellar populations and gas properties of the Seyfert 2 galaxy NGC\,232. Our main conclusions are:

\begin{itemize}
    \item Based on emission-line ratio diagnostic diagrams, the gas emission in NGC\,232 displays components attributed to both star formation and AGN activity. The AGN-dominant region is revealed to be an ionization cone, oriented with its southeastern side facing toward us, seen in front the galaxy's disc plane, while the northwestern side points in the opposite direction, being obscured by the galaxy's disc.

    \item The gas kinematics present two components: a disc component which originate narrow line components ($\sigma\lesssim 150$\,km\,s$^{-1}$), following a similar rotation pattern as the stars with the line of nodes seen along the position angle $\Psi_0\approx30^\circ$; and a biconical outflow, seen as broad ($\sigma\gtrsim 150$\,km\,s$^{-1}$) and mostly blueshifted components in the emission line profiles. 
   \item The stellar velocity field is well reproduced by a rotating disc model, as for the gas, residuals of up to 60\,km\,s$^{-1}$ are seen between the observed and modeled velocities, whitin the inner 1.5 kpc radius. These non-circular motions are likely produced by the interaction of outflowing gas with the gas in the disc or by a recent merger.

    \item We estimate a maximum outflow rate of 1.26$\,\pm$0.16\,M$_{\odot}$\,yr$^{-1}$ observed at $\sim$560 pc from the nucleus. The maximum kinetic power of the outflow is  $3.4\pm0.4\times10^{41}$ erg\,s$^{-1}$, which corresponds to $\sim$0.003\,$L_{AGN}$. The energy released seems to be enough to quench star formation within the cone, as suggested by the lower star formation rates in this region.

    \item Old ($t >2\,Gyr$) stellar populations dominates the mass and light fractions. The maximum emission contribution of the old stellar populations is seen at the nucleus, decreasing with its distance. On the other hand, young and intermediate age stellar populations present increasing contribution profiles with the distance to the nucleus. Along with an increasing gradient of the stellar metallicity,  these results are consistent with an inside-out star-formation scenario in NGC\,232.
    
    \item  The $SFR_{\star}$ map reveals a circumnuclear star forming ring with $\sim\,440 \,pc$ radius, which could be indicative that both the AGN and the circumnuclear star formation might have been triggered by the same gas reservoir. Alternatively, the AGN could be triggered by winds from young stars in the ring. 

    \item Regions where the gas emission is produced by photoionization by the AGN and locations where the emission is dominated by young stars show similar oxygen abundances. The mean values are $\langle 12 + \log({\rm O/H})\rangle=8.80\pm0.05$ for the SF dominated spaxels and  $\langle 12 + \log({\rm O/H})\rangle=8.73\pm0.09$ for the AGN dominated region. 

\end{itemize}
\section*{ACKNOWLEDGEMENTS}

We would like to thank the referee for their valuable comments that improved our final manuscript. H.C.P.S. thankS the financial support from Coordena\c c\~ao de Aperfei\c coamento de Pessoal de N\'ivel Superior - Brasil (CAPES) - Finance Code 001. R.A.R. acknowledges the support from Conselho Nacional de Desenvolvimento Cient\'ifico e Tecnol\'ogico (CNPq; Proj. 400944/2023-5 \& 404238/2021-1)  and Funda\c c\~ao de Amparo \`a pesquisa do Estado do Rio Grande do Sul (FAPERGS; Proj. 21/2551-0002018-0). O.L.D. Thanks to Universal FAPESP/SECTI N.012/2022 (proj. 5017/2022).  Based on data obtained from the ESO Science Archive Facility.

\section*{Data Availability}
The data used in this work are available at the ESO science archive under the project code  095.D-0091(B)



\bibliographystyle{mnras}
\bibliography{paper.bbl} 




\section{Emission line flux maps}


Figure~\ref{fig:ap_flux} shows the flux maps for the narrow (left panels) and broad (right panels) components for the following emission lines: [O\,{\sc iii}]$\lambda5007$, [O\,{\sc i}]$\lambda6300$, H$\alpha$, [N\,{\sc ii}]$\lambda6584$ and [N\,{\sc ii}]$\lambda6716$.

\appendix
\begin{figure*}
    \includegraphics[width=.75\textwidth]{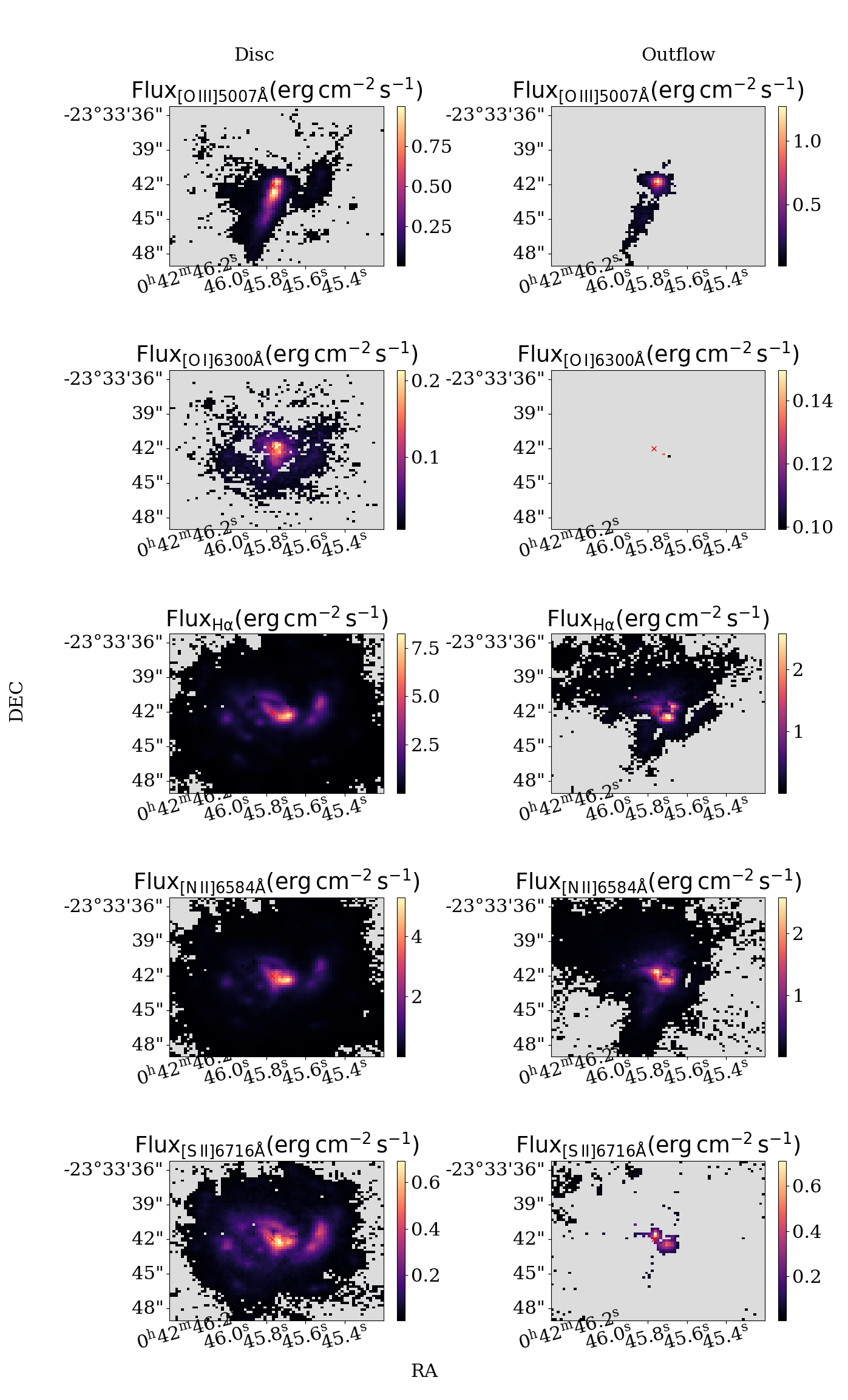}
    \caption{Flux maps for the optical emission lines in units of $10^{-16}\,{\rm erg\,cm^{-2}\,s^{-1}}$. The red cross marks the peak of the continuum.}
    \label{fig:ap_flux}
\end{figure*}


\bsp	
\label{lastpage}
\end{document}